# Enhancing Chemical Stability and Photovoltaic Properties of Highly Efficient Nonfullerene Acceptors by Chalcogen Substitution: Insights from Quantum Chemical Calculations

*Leandro Benatto,[a,b\*] João Paulo A. Souza,[a] Matheus F. F. das Neves,[a] Lucimara S. Roman[a], Rodrigo B. Capaz,[b,c] Graziâni Candiotto[b\*] and Marlus Koehler[a]*

[a] *Department of Physics, Federal University of Paraná, 81531-980, Curitiba-PR, Brazil*

[b] *Institute of Physics, Federal University of Rio de Janeiro, 21941-909, Rio de Janeiro-RJ, Brazil*

[c] *Brazilian Nanotechnology National Laboratory (LNNano), Brazilian Center for Research in Energy and Materials (CNPEM), Campinas, São Paulo, 13083-100, Brazil*

\*Corresponding authors: lb08@fisica.ufpr.br ; gcandiotto@iq.ufrj.br

Abstract

The chemical stability of nonfullerene acceptor (NFA) is the Achilles' heel of the research on state-of-the-art organic solar cells (OSC). The fragility of the NFA is essentially due to the weak bond that links the central donor core of the molecules with their acceptor moieties at the edges. Here we proposed the replacement of thiophene at the outer-core position of traditional NFAs for tellurophene, a hitherto unexplored modification. Since tellurium is a distinctive element among chalcogens, the basic features of Te compounds cannot be deduced straightforwardly from the properties of their lighter analogues, S and Se. The modeled Te-based NFAs presented interesting features like stronger intra- and intermolecular interactions induced by a distinctive secondary bond effect between the end acceptor moiety and the outer chalcogen atom. This design strategy resulted in stiffer molecules with red-shifted absorption spectra and less susceptible to degradation verified through stress tests and vibrational spectra analysis. Besides that, a weakened exciton binding energy has been found, opening the possibility of blends with a lower driving force. Our results shed light on several aspects of selenation and telluration of traditional NFAs, providing valuable insights into the possible consequences for OSCs applications.

## 1. Introduction

Nonfullerene acceptors (NFAs) are under intense research in recent years, leading to significant improvements in the performance of organic solar cells (OSCs).[1] With the delicate design and synthesis of A-D-A-type acceptors such as ITIC and A-DA′D-A-type acceptors such as Y6 (D is the electron donor molecular unit and A is the electron acceptor molecular unit), the efficiency of OSCs has reached around 19%.[2] Despite the increase in efficiency, the relatively short device lifetime and the low photostability of NFAs is still an issue to be solved.[3] Luke *et al*.[4] find that a critical feature that causes NFAs degradation is a twisted conformation, facilitating the photoisomerization, fragmentation and photo-oxidation. In addition, it was found that the C–C bridge bond between "A" and "D" units of NFAs is particularly fragile against many kinds of external stresses.[5] More planar and rigid NFAs are found to be less susceptible to photodegradation.[6] Other strategies have been proposed to increase the photostability of NFAs like, end-group fluorination that increases molecular rigidity[7] and less bulk side chains to promote dense packing of molecules.[8] Besides that, the replacement in the chemical structure of thiophene for heavier chalcogenophenes (selenophene and tellurophene; chalcogens are the chemical elements in group 16) proved to be an interesting way to increase the crystallinity of donor conjugated copolymers,[9] perylene diimide (PDI) acceptors,[10] and more recently polymer acceptors.[11–13]

The search for NFAs with high photostability has been intense, generating interesting molecular insights like volume-conserving photoisomerization acting as a surrogate towards subsequent photochemical reaction.[8] Another important discovery reported by Labanti *et al*.[14] lead to OSCs with high operational stability by replacing sulfur (S) atoms for selenium (Se) atoms at outer-core positions of ITIC-based NFAs. The incorporation of selenophene moieties in NFAs is important to increases intra- and intermolecular interactions.[15] This strategy is also useful to tune energy levels and enhance light-harvesting ability in visible and near-infrared (NIR) regions, increasing important device parameters like the open-circuit voltage ($V_{OC}$), photocurrent density ($J_{SC}$), and power conversion efficiency (PCE).[3,16–19]

The introduction of heavier and larger atoms at outer-core positions can also impact the device performance by inducing a higher density of triplet excitons. It is known that the addition of heavy atoms in the molecular structure can increase spin-orbit coupling (SOC) facilitating the generation of triplet states by intersystem crossing (ISC).[20] In principle, the forbidden nature of triplet recombination is interesting to promote longer exciton lifetimes and diffusion lengths, which can help raise the device's efficiency.[21] However, triplet excitons have higher binding energies compared to singlet excitons,[22] which makes generating free charges from them a greater challenge.[23] It is imperative to emphasize that, given the potential for energy loss within the triplet states, a better understanding on the mechanisms governing triplet dynamics in NFAs should be established.[24,25]

The change of the molecular electronic structure induced by chemical modification can also impact dipole moment ($\mu$) and quadrupole moment ($Q$). The fine tune of $\mu$ can regulate import parameters like the nanoscale morphology,[26] driving force[27] and exciton dissociation[28] in donor/acceptor blends. There are reports that large quadrupolar moments destabilizes the CT state facilitating the exciton dissociation.[29,30] Its impact on the energy levels at the interfaces between two organic semiconductors is related to charge-quadrupole interaction.[31,32] Besides that, the charge-quadrupole interaction induces morphology control facilitating the formation of more ordered and compact molecular packing.[33] The search for NFAs with high polarization, for example with Se and Te atoms, can raise these interesting effects that increase the efficiency of OSCs.

Possibly due to the scarcity of tellurium,[34] the replacement of thiophene for tellurophene has been less explored compared to selenophene in NFAs.[35] This motivated us to study the influence of these heavy atoms on NFAs through quantum chemical calculations. This approach makes it possible to obtain relevant molecular information from NFAs already produced in addition to studying some structures not yet synthesized. Significant differences related to the substitution of S by Se or Te can be evidenced by density functional theory (DFT)[36] and time-dependent DFT (TD-DFT)[37] calculations, leading to a deeper understanding of the correlation between electronic properties, chemical structural stability, and photovoltaic response.[38–45]

In this study, the effects of selenation and telluration of A-D-A and A-DA′D-A-type NFAs were theoretically investigated by DFT/TDDFT. Two trios of NFAs were chosen. We first choose two consolidated NFA from the literature to replace two S atoms for two Se or Te atoms. Based on encouraging reports in the literature,[46] we decided to use the molecules Y6 (2,2'-((2Z, 2'Z)-((12,13-bis(2-ethylhexyl)-3,9-diundecyl-12,13-dihydro-[1,2,5]thiadiazolo[3,4-e]thieno[2″,3''':4',5']thieno[2',3':4,5]pyrrolo[3,2-g]thieno[2',3':4,5]thieno[3,2-b]indole-2,10-diyl)bis(methanylylidene))bis(5,6-difluoro-3-oxo-2,3-dihydro-1H-indene-2,1-diylidene))dimalononitrile)[47] and ITIC (3,9-bis(2-methylene-(3-(1,1-dicyanomethylene)-indanone))-5,5,11,11-tetrakis(4-hexylphenyl)-dithieno[2,3-d:2',3'-d']-s-indaceno[1,2-b:5,6-b']dithiophene).[48] Here, the replacement of S for Se or Te were performed at outer-core positions following the reports of good stability by Labanti *et al.*,[14] and good photovoltaic parameters reported by Zhang *et al.*[49]. The chemical structures of the molecules are presented in Figure 1. Note that Y6-based molecules present a banana-shaped structure[50] and ITIC-based molecules a S-shaped structure.[51,52] Among the molecules simulated in this work in which S atoms were replaced, there are reports in the literature of ITIC-2Se[14] and Y6-2Se.[53,54] Therefore, to our knowledge, the Te-containing molecules studied here can be considered not synthesized so far. Yet tellurium-containing aromatics, tellurophene-based small molecules and polymers have been reported in the literature.[55,56] Recently, dibenzotellurophene (DBTe) and larger conjugated molecules have been synthesized as promising materials for different applications including solar cells.[57] In addition, more efficient synthetic methods to produce DBTe were proposed. Since DBTe is the fundamental unit for large tellurium-containing aromatics, one can anticipate an increasing interest in the use of Te-based materials for optoelectronic applications. Yet the properties of larger π-extended tellurium-containing conjugated systems are still poorly understood,[55] as their S and Se counterparts are better known organic semiconductors. Hence, more research is needed to explore the potentials of Te-based conjugated materials, especially because Tellurium is a unique element among chalcogens, and the basic features of Te compounds cannot be deduced straightforwardly from the properties of their lighter analogues.[56,57]

From a theoretical point of view, given the few reports in the literature,[10,15] in-depth simulations (beyond optimized geometries by DFT at a low theory level) of NFAs containing heavy atoms in the chemical structure is challenging because it involves a high computational cost. Although computationally expensive, a long-range functional and larger basis set was used here in the simulations to carry out an in-depth analysis of the structural modifications of the NFAs caused by the addition of Se or Te (see the Supporting Information for computational details). Our results indicate that replacing S atoms for Se atoms (and especially for Te atoms) at outer-core positions of NFAs can enhance the chemical stability of those acceptors. A higher resistance toward external stresses is obtained by strengthening the C–C and C=C bonds between "A" and "D" units of the NFA and a stronger intramolecular electrostatic interactions between the oxygen lone pairs and the chalcogen atoms – witch acquires increasingly large positive charges as the atomic size increases – located at the terminal 2-(3-oxo-2,3-dihydroinden-1-ylidene)malononitrile (INCN) group and the central core of the NFA. The deeper understanding of this effect can reveal new approaches to mitigate the degradation mechanism of state-of-the-art OSC.

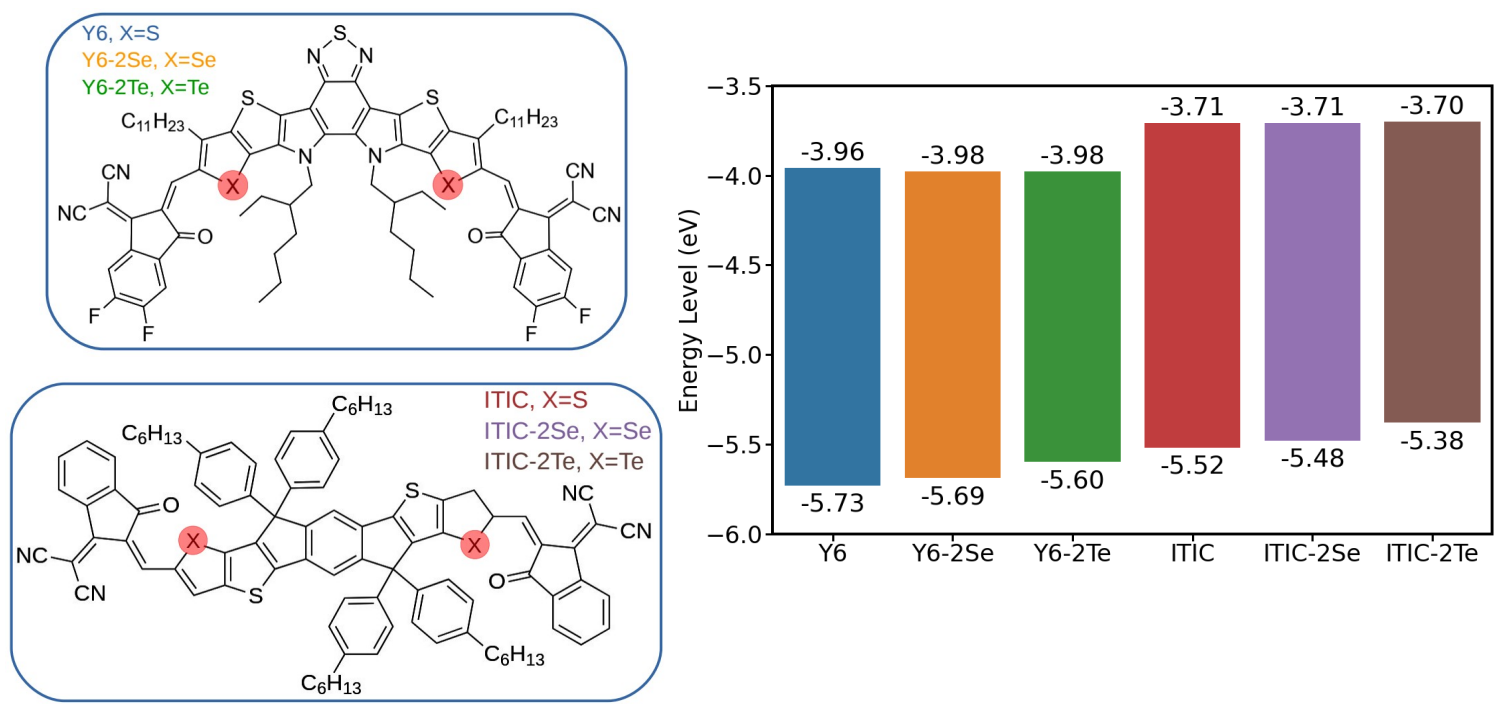

**Figure 1**: Molecular structures of the acceptors simulated in this study and the calculated energy-level diagram.

## 2. Results and Discussions

### 2.1 Structural modifications

The chemical structures of the studied acceptor molecules are shown in Figure S1. The calculated energies of the highest occupied molecular orbital (HOMO) and lowest unoccupied molecular orbital (LUMO) are presented in Figure 1. The theoretical results are close to those estimated from electrochemical measurements with cyclic voltammetry available in the literature (Tables S1 and S2). There is a systematic decrease in the HOMO-LUMO gap of molecules when replacing S by Se or Te, an important modification that enhance their light-harvesting ability. The gap reduction is mainly due a significant variation of the HOMO energy, induced by a higher degree of electronic delocalization (see discussion below). In addition, a red-shifted absorption was found for tellurophene-based materials compared to sulfur and selenium analogues[55] which is confirmed by the TDDFT simulations shown in Figures S2 and S3. Although efficient free charge generation has already been obtained in several D/A blends, a strategy to further increase the $V_{OC}$ and PCE of OSCs involves the combination of materials with the minimal driving force to promote the exciton dissociation.[58] Therefore, a fine tunning of the energy levels is important to improve the photovoltaic device.[59] The changes observed in energy levels are related to the structural modifications and also to variations in charge distribution as will be seen below.

Important structural variations resulting from selenation and telluration of Y6 and ITIC are outlined in Figure 2. The parameters chosen to probe structural changes are the molecular length, $l$, the bond lengths X−C1, C1−C2, C2=C3 and X…O (X=S, Se or Te), the angle X−C1−C2 and the dihedral angle X−C1−C2−O (see Figure 2a). Note that the bond length between central fused-ring core and acceptor end-group is defined by the C1−C2 distance. The dihedral angle between central fused-ring core and acceptor end-group is defined by X−C1−C2−O. An additional important parameter, only for the banana-shaped Y6-based molecules, is the opening angle, $\alpha$. The values of these structural parameters are shown in Table 1. Note the increase in magnitude of $l$, $\alpha$ and X−C1 with selenation and telluration of the molecules. The results shows that the molecules become more linear, with a decrease in the banana-shaped structure of Y6-based molecules and a decrease in the S-shaped structure of ITIC-based molecules. Notably, the variations are more pronounced with telluration. We anticipate that this trend will be maintained for other properties analyzed here. For instance, due to the larger size of the Te atom, the longest X−C1 bond length (2.113 Å) is obtained for the Tellurium substitution. Our calculated length is comparable to other estimates of the Te−C bond distance obtained theoretically (2.100 Å)[55] and experimentally (2.055 Å).[60]

Another noticeable structural variation is the decrease of X…O distance with selenation and telluration, i.e., there is an approximation between the oxygen of the acceptor end-group with X that is part of the central fused-ring core of the molecules. This approximation, which is essentially produced by an increase in O polarization with the introduction of Se or Te atoms (discussion below), causes a slight increase in the HOMO−LUMO overlap, $\Theta_{H-L}$, as can be seen in Figure S4. Note that the HOMO is localized mainly in the central fused-ring core of the molecules, whereas the LUMO is localized mainly in the acceptor end-group.

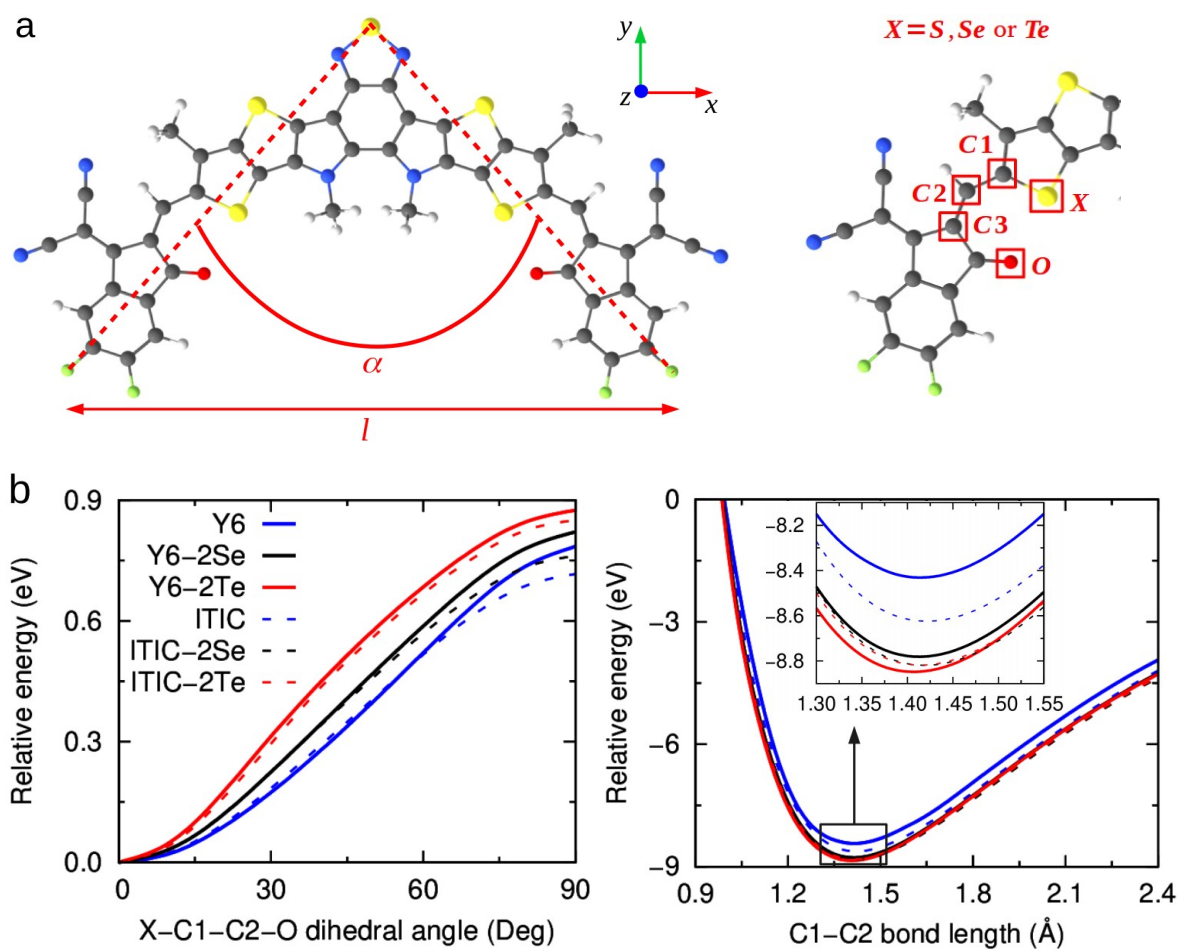

**Figure 2**: (a) Scheme of the most relevant structural parameters analyzed in this study. (b) *left*: Relative molecular potential energy *vs* the dihedral angle between central fused-ring core and acceptor end-group, *right*: Relative molecular potential energy *vs* the bond length between central fused-ring core and acceptor end-group.

**Table 1**: Results of the most relevant structural parameters analyzed in this study for each simulated molecule. Bond lengths (X−C1, X⋯O, C1−C2, and C2=C3), bond length alternation (BLA) between C1−C2 and C2=C3, and molecular length (*l*) in Å. X−C1−C2 angle, X−C1−C2−O dihedral angle and opening angle (α) in degrees. C1−C2 and C2=C3 bonding energy in eV.

| Molecules | X−C1 | X⋯O | C1−C2 | C2=C3 | BLA | X−C1−C2 | X−C1−C2−O | *l* | α | $E_{bond}^{C1-C2}$ | $E_{bond}^{C2=C3}$ |
|---|---|---|---|---|---|---|---|---|---|---|---|
| Y6 | 1.77 | 2.70 | 1.422 | 1.373 | 0.049 | 127.19 | 1.10 | 22.99 | 81.56 | −8.43 | -7.94 |
| Y6-2Se | 1.90 | 2.69 | 1.420 | 1.375 | 0.045 | 127.64 | 1.14 | 24.28 | 86.70 | −8.78 | -8.01 |
| Y6-2Te | 2.12 | 2.64 | 1.415 | 1.380 | 0.035 | 127.74 | 0.57 | 25.53 | 91.83 | −8.85 | -9.37 |
| ITIC | 1.76 | 2.72 | 1.430 | 1.368 | 0.062 | 129.19 | 0.38 | 28.99 | - | −8.62 | -7.97 |
| ITIC-2Se | 1.90 | 2.71 | 1.428 | 1.370 | 0.058 | 129.87 | 0.87 | 30.34 | - | −8.82 | -8.06 |
| ITIC-2Te | 2.11 | 2.68 | 1.424 | 1.375 | 0.049 | 130.57 | 0.94 | 30.72 | - | −8.82 | -8.22 |

Regarding the dihedral angle between these two chemical groups, the results show extremely low values, practically equal to zero for all molecules, with no change due to the addition of Se or Te atoms. It is known that the high planarity of ITIC is due to the O⋯S intramolecular noncovalent interaction of the central adjacent rings.[61] The results obtained here show that this intramolecular interaction is increased by replacing S atoms by Se or Te. This effect is demonstrated by calculating the relative molecular potential energy *vs* the dihedral angle between central fused-ring core and acceptor end-group, or *vs* the bond length between central fused-ring core and acceptor end-group. See on the left side of Figure 2b that the rotational energy barrier increases with the inclusion of Se or Te atoms. Likewise, there is a systematic reduction of the C1–C2 bond length with the replacement of S by increasingly heavier chalcogens (see Table 1). The shortening of this length indicates that the C–C linkage bond between "A" and "D" units of NFAs is strengthened with these substitutions. Looking at the length of the C2=C3 bond, a very slight increase is observed. From the length of these two bonds that connect the D and A regions of the molecules, we calculate the bond length alternation (BLA), which is the difference between them. The continuous decrease in BLA in this region demonstrates an increase in the degree of hybridization of these bonds and reinforces the idea that molecular stiffening is occurring. Signatures of this strengthening can also be observed in the vibrational spectra, as discussed in Section 2.3. On the right side of Figure 2b, it can be observed that the C1–C2 bonding energy $\left(E_{bond}^{C1-C2}\right)$ is increased (indicated by a deeper potential well) for molecules containing Se or Te, showing that the D and A groups of the molecules are more strongly bound. The C2=C3 bonding energy $\left(E_{bond}^{C2=C3}\right)$ also increases, as can be seen in Table 1. It is well recognized that avoid side reaction of exocyclic vinyl group is decisive to improve the intrinsic stability of A–D–A NFAs.[8] Hence all those findings suggest that the modified molecules are more rigid, which might possibly increase the photostability of these materials.

The structural results described above suggests that there is an intramolecular secondary bond between the X atom of the donor and the O atom of the acceptor INCN end-groups. Secondary bonds are characterized by interatomic distances that are longer than covalent single bonds, but shorter than the sum of van der Waals radii.[56] Considering that the van der Waals radii of O, S, Se and Te are 1.55, 1.8, 1.9, and 2.1 Å,[62] respectively, the X⋯O distances of Table 1 fulfills this criterion. Moreover, secondary bonds are also called σ-hole interactions[56] because they are formed by the interaction between a positively charged site (σ-hole) and a negatively charged site (containing a lone pair) that are located in different atoms (Table 2 below shows the polarization of the X and O atoms). This effect can be clearly observed in Figure 3, wherein there is a notable increase in the region of overlap between the van der Waals radii for the Te-functionalized molecule. This kind of non-covalent bond essentially derives from a $n^2$(D) → σ*(E–Y) interaction in which the lone pair of a donor atom D partially mix with the σ* orbital of the bond between a heavy atom (E) and a more electronegative atom (Y).[56] Since the polarizability tends to increase with increasing atomic number, the energy difference between the σ(E–Y) and σ*(E–Y) orbitals decreases as E gets heavier. This ends up in stronger σ-hole interactions for tellurium compared to those of selenium and sulfur. A strengthened intramolecular secondary bond joining the end acceptor groups with the central donor groups of the tellurium substituted NFAs can favor the chemical stability of those molecules.

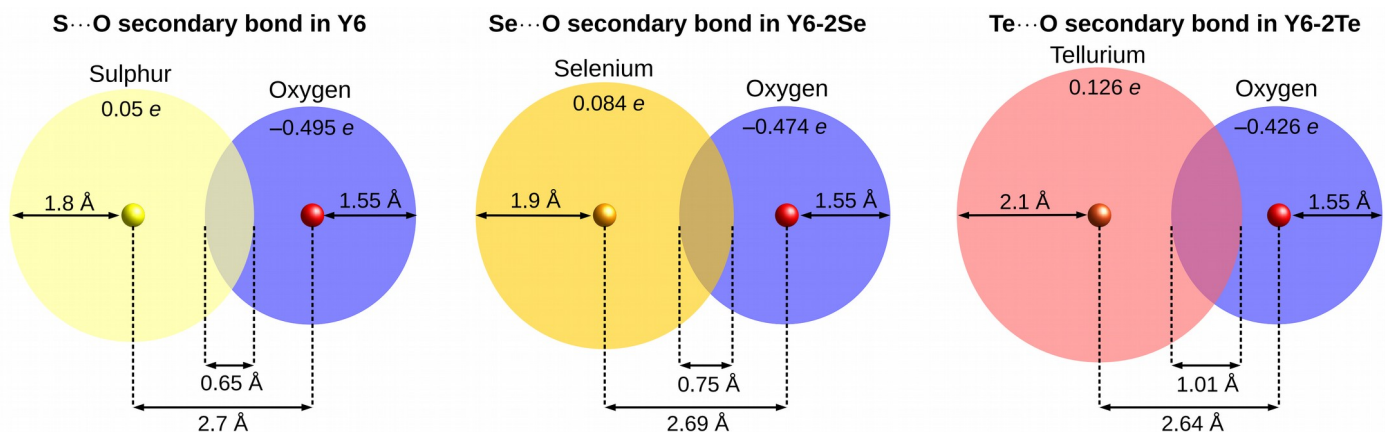

**Figure 3**: Illustrative scheme of secondary chemical bond strengthening for the studied Y6-derived molecules. In the figure we present the van der Waals radius of the atoms, the interatomic distance, the overlapping region, and the atomic partial charges.

## 2.2 Charge distributions

For all considered molecules, the restrained electrostatic potential (RESP)[63] fitting method was employed to obtain accurate atomic partial charges, which was calculated using the Multiwfn 3.8 package.[64] The partial atomic charges were analyzed and show an increase in local polarization with the introduction of Se or Te atoms. We will focus first on the X and O atoms highlighted in red in Figure 2. As can be seen in Table 2, the partial charge of S atom is near zero, much smaller than Se and Te. The highest values were obtained for Te atoms probably because tellurium has a lower electronegativity than sulfur or selenium. It was also found that oxygen atom became less negative in molecules with Se and Te. Therefore, the heavier chalcogens interact more strongly with oxygen, which increases local polarization. Figure 4 illustrates the contours maps of the electrostatic potential (ESP) of the six acceptor molecules. The ESP distribution has proven to be a highly effective tool for investigating noncovalent interactions with great success.[65,66] The electrostatic potentials are visually represented by red (positive) and blue (negative) colors. To provide a closer examination of the X⋯O interaction, a zoomed-in region has been included. Compared to S⋯O interaction the results revealed a notable reduction in the blank region (zero ESP region) between Se⋯O atoms and mainly between Te⋯O atoms, resulting in the increase of an electrostatic interaction. Specifically, the increase of ESP in the region between the atoms promoted the strengthening of secondary bonds, leading to enhanced interatomic forces. This result further reinforces the notion that the chemical changes proposed in this study contribute to enhanced molecular stability.

**Table 2**: Partial charge ($q$, in unity of electron) of the atoms X and O, magnitude of the molecular dipole moment in ground state and excited state ($\mu_g$ and $\mu_e$, in Debye), change in dipole moment ($\Delta\mu_{ge}$), molecular quadrupole moment ($Q$, in Debye Å), internal charge transfer (ICT, in unity of electron) and intramolecular reorganization energy ($\lambda_{int}$, in eV).

| Molecules | $q_X$ | $q_O$ | $\mu_g$ | $\mu_e$ | $\Delta\mu_{ge}$ | $Q_\pi$ | ICT | $\lambda_{int}$ |
|---|---|---|---|---|---|---|---|---|
| Y6 | 0.050 | –0.495 | 1.83 | 0.82 | 1.03 | 77.19 | 0.711 | 0.265 |
| Y6-2Se | 0.084 | –0.474 | 3.05 | 2.10 | 0.97 | 75.15 | 0.704 | 0.257 |
| Y6-2Te | 0.126 | –0.426 | 4.95 | 4.08 | 0.92 | 70.88 | 0.641 | 0.251 |
| ITIC | 0.033 | –0.427 | 0.21 | 0.49 | 0.28 | 50.22 | 0.486 | 0.254 |
| ITIC-2Se | 0.075 | –0.408 | 0.24 | 0.67 | 0.43 | 48.83 | 0.468 | 0.265 |
| ITIC-2Te | 0.130 | –0.368 | 0.56 | 0.91 | 0.35 | 44.99 | 0.437 | 0.255 |

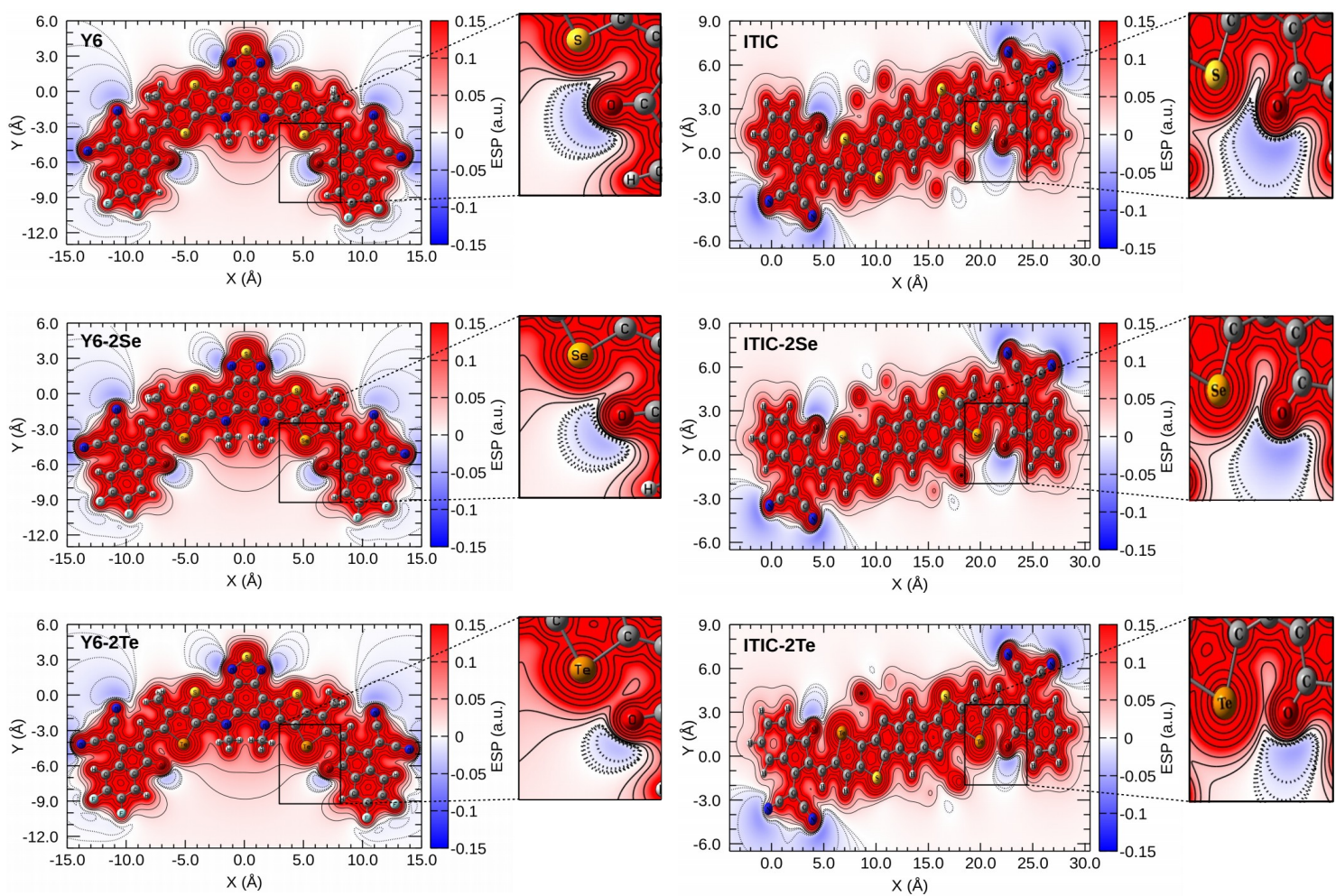

Figure 4 - Contours maps of the electrostatic potential (ESP) of investigated acceptor molecules.

The growing local polarization ($q_x$ partial charge) with the introduction of Se or Te atoms has a pronounced impact on the molecular electric dipole moment in the ground state geometry ($\boldsymbol{\mu}_g$), as clearly demonstrated in Table 2. The dipole moment vector of Y6-based acceptors is located at the center of the acceptor with a preferential orientation parallel to the molecule's plane (y-direction, see Figure S5). The addition of Se and Te atoms to the molecular structure intensifies the charge imbalance along the molecule's plane which further orients the dipole moment along y. The intensification of the asymmetry of the charge distribution in the y-direction also impacts the component $Q_{yy}$ of the quadrupole moment, as can be seen in Figure S5. The variation of $\boldsymbol{\mu}_g$ is especially high for the Y6-based acceptors due to their banana-shaped structure suitable for the emergence of the dipole moment as the local polarization increases. In ITIC-based acceptors, on the other hand, $\boldsymbol{\mu}_g$ is close to zero due to the symmetry of their molecular structure, so that the increase in local polarization has a smaller effect in $\boldsymbol{\mu}_g$. The fine tune of $\boldsymbol{\mu}_g$ can impact positively the nanoscale morphology, driving force and exciton dissociation.[26,27]

It was also verified that the heavier chalcogens slight decreases the internal charge transfer (ICT) between donor and acceptor moieties (see Table 2). This effect could be related to the trapping of negative charges in the region where the heavier chalcogen is inserted, making it difficult for then to migrate to the acceptor end-group. Therefore, the heavier chalcogens slight reduces the concentration of negative charge in the acceptor end-groups. Comparing the two groups of simulated molecules it is important to note that Y6-based acceptors that possesses fluorine atoms at the end-groups presents a higher ICT than ITIC-based acceptors. In addition, the dipole moment change between the ground state and excited state, $\Delta\mu_{ge}=|\boldsymbol{\mu}_g - \boldsymbol{\mu}_e|$, is much superior for Y6-based acceptors suggesting an easier free charge generation for OSCs applications.[67]

Another relevant parameter to be analyzed is the quadrupole moment of the molecules. The intermolecular quadrupole-quadrupole interactions can facilitates the intermolecular packing promoting morphology optimization.[33] Besides that, quadrupole interactions at the donor–acceptor interface can assist the free charge generation due to electrostatic potential destabilization of the CT state.[68] The results showed higher values for $Q_{zz}$ (or $Q_\pi$, the quadrupole component along the π-π stacking direction) for Y6-based acceptors compared to ITIC-based acceptors, which can be attributed to the presence of flour atoms at the end-groups.[33] The values for Y6-based acceptors are around 75 Debye Å, which is considered excellent to provide a balance between efficient exciton dissociation and open circuit voltage losses.[30] However, there is a small decrease in $Q_\pi$ with selenation and telluration. This slight reduction may be attributed to a more symmetric charge distribution along the π-π-stacking direction.[31]

Considering the intramolecular reorganization energy for electron transfer ($\lambda_{int}$),[69] slightly lower values were found with selenation or telluration of Y6. This effect was not observed for ITIC-based acceptors, since the values of $\lambda_{int}$ did not show a clear trend. Reduced values of the reorganization energy for electron transfer are important to lower the energy barrier for electron transfer,[70] facilitating their transport along the material. Despite the small difference in $\lambda_{int}$ between molecules of the same type, there are sound evidences that charge transport is improved in relevant materials with the addition of Se atoms.[35] This indicates that packing effects must play an important role in the electron mobility of the acceptor molecules considered here.

**2.3 Simulated Raman spectra**

Raman spectroscopy (RS) is a very useful technique to characterize materials. It has been previously applied to study the stability of organic photovoltaic materials, yielding valuable insights about photo-induced chemical variations and their origins.[71] These important characteristics of the RS method motived us to simulate the Raman spectra of the novel NFAs considered in Figure 1.

Simulated Raman spectra for Y6 and ITIC in Figure 5 are similar to experimentally reported in the literature,[71,72] since the main characteristic bands of those molecules are present in the simulations. The graphs are shown for the range between 1150 and 1550 cm$^{-1}$, and the bands outside this interval can be accessed in Figure S6 in the SI file. For instance, Alkene band is computed in the regions of 1550, while peaks related to phenyl and carbonyl groups are found at 1600 and 1705 cm$^{-1}$, respectively.[71] For both, ITIC and Y6, the region of 1070 and 1400 cm$^{-1}$ corresponds to C–N stretching vibrations. In addition, the calculated Raman shifts at 614 and 720 cm$^{-1}$ can be associated to C–S stretching vibrations.[73] Bands associated to thiophene at 830 and 770 cm$^{-1}$ due to ring deformation and at 1225 and 1460 cm$^{-1}$ due to C=C can also be found in calculated Y6 and ITIC Raman structures.[74]

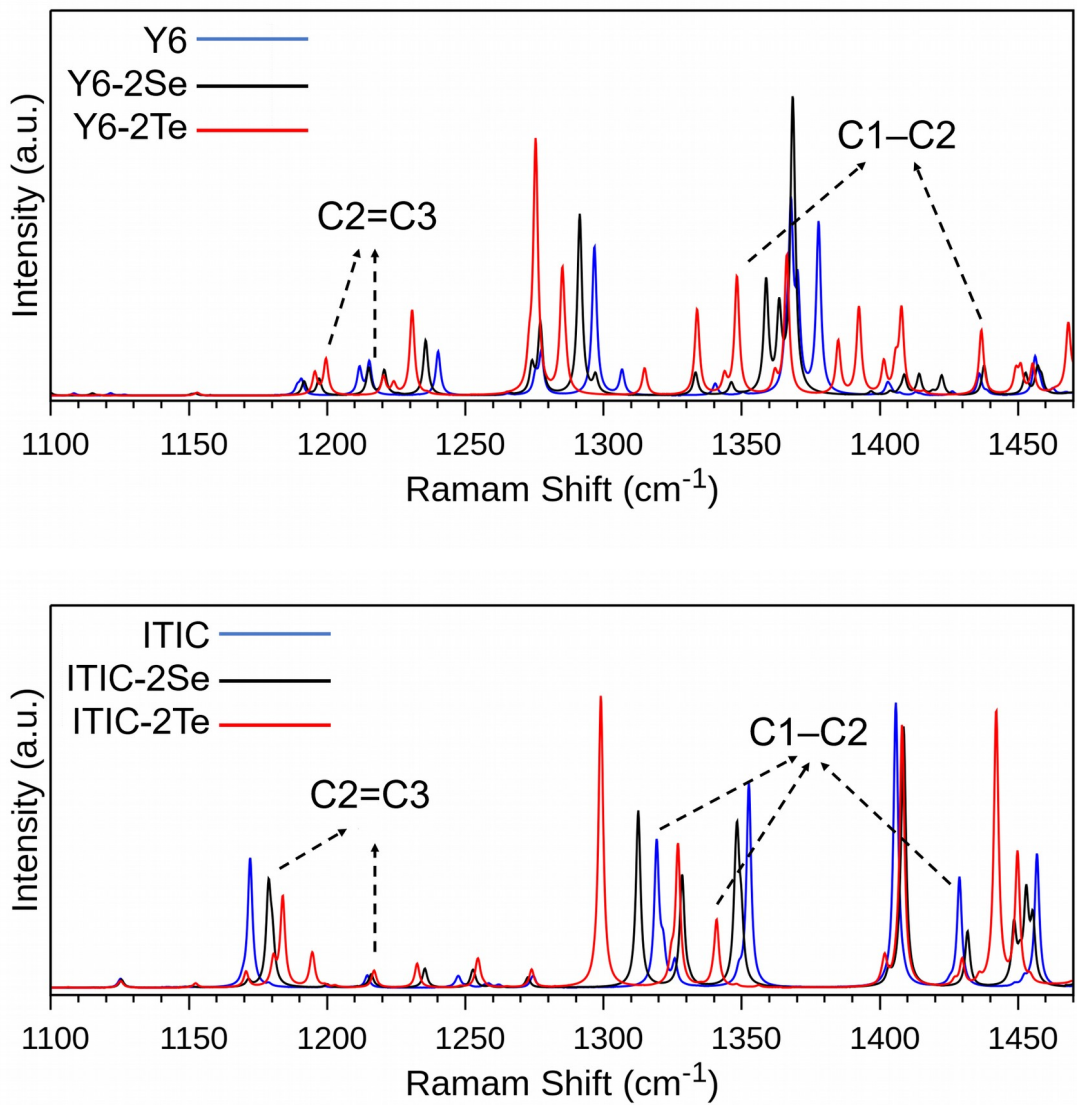

**Figure 5**: Simulated Raman spectra showcasing the vibrational signatures of the acceptor molecules. *top*: Y6-based molecules, *bottom*: ITIC-based molecules. Inset: the arrows indicate the blueshift of the indicated chemical bond peaks produced by selenation and telluration.

For Y6 molecule, the peak relative to the β-CNC bond appears in the region around 370 cm$^{-1}$. Bands with the frequencies of 1260 and 1364 cm$^{-1}$ are referent to $C_\alpha$–$C_\alpha$ inter-ring stretching and $C_\beta$–$C_\beta$ stretching, respectively.[75,76] The first is more accentuated in Y6 structure, while the second can be observed in both. Bands in the region at 1450 cm$^{-1}$ are associated to $C_\alpha$–$C_\beta$ symmetric stretching, while at 1515 and 1560 cm$^{-1}$ are associated to $C_\alpha$–$C_\beta$ asymmetric modes.[77–79] These bands are present in both structures with maximum intensity registered for molecules with selenophene and tellurophene, respectively. It is also possible to note that these bands are red-shifted as S is substituted by Se and Te, with more shift associated to tellurophene structure.

For both kind of molecules, the thiophene bands such as at 1425 cm$^{-1}$ presents lower intensities as Se and Te are included. This effect can be associated to the addition of selenophene and tellurophene rings, respectively, thereby decreasing the intensity of the thiophene bands.

For ITIC the band in the region around 1475 cm$^{-1}$ presents a red shift for both structures as S is substitute by Se and Te. It can be associated to the lower availability of thiophene groups and also a higher delocalization of π-electrons across selenophene and tellurophene units in the conjugated chains.[80]

The addition of selenophene structure in ITIC makes the bands due to carbon and selenium appears in the spectrum. The region in 459 and 230 cm$^{-1}$ are due to Se cyclic and chain structures, respectively, and at 580 – 590 cm$^{-1}$ are characteristic of Se–C vibration modes.[81] For Y6-2Se structure these bands are present at 287 and 1100 cm$^{-1}$, corresponding to Se=C[82] and Se–C[83] respectively.

Bands in low-frequency regions (*e.g.*, lower than 600 cm$^{-1}$) are characteristics of molecules with heavier atoms. This explains the appearance and increase of bands in the low-frequency interval of the spectrum as Se and Te substitutes S in the NFAs. For example: the band at 565 cm$^{-1}$ is associated to C–Te–C,[84] and at 300 cm$^{-1}$ is a characteristic peak associated to the Te presence.[85] Both structures are clearly present in the Y6-2Te and ITIC-2Te spectra.

Besides the detection of new low-frequency bands, the presence of selenophene and tellurophene also produces a small blueshift in some peaks observed for the unsubstituted molecules. Observing the peaks of thiophene C=C in the region between 1150 and 1225 cm$^{-1}$ in ITIC spectra it is clearly observed this shift to higher frequencies as the heavier atoms are placed in the molecule structure. The same is observed in the region between 1300 and 1350 cm$^{-1}$ that is one of the regions associated to the C1–C2 bond expressed in figure 2a. Another region that corresponds to this same carbon bond is around 1425 cm$^{-1}$ and one observed feature is the appearance of an additional shoulder in both spectra, indicating new vibrational modes introduced by selenation and telluration. For Y6 spectra the shift for higher frequencies is observed for the same bond groups C=C and C1–C2, corresponding to the regions around 1170 - 1230 cm$^{-1}$ and 1350, 1400-1450 cm$^{-1}$, respectively. These blueshifts can be associated to the addition of the heavy chalcogens Se and Te. From figures 2a and 4 it can be seen that both bonds interact with oxygen atom and the different positions for the vibrational modes indicates a shortening in the bond length and a stiffen in this region of the molecules as a result of the intensified intramolecular secondary bond interactions. This result is interesting because it confirms a strengthening of the molecular region most sensitive to degradation, which is in accordance with our previous discussions.

## 2.4 Excited states

Figure 6a shows the calculated singlet ($E_{b,S}$) and triplet ($E_{b,T}$) exciton binding energies of the acceptors. One can see that there is a noticeable weakening of $E_{b,S(T)}$ with selenation and telluration. This effect can have a strong influence on the photovoltaic response of these materials since the reduction of the exciton binding energy opens the possibility of using D/A blends with a lower driving force, an excellent strategy to increase the $V_{OC}$ of OSCs.[86] The reduction in excitonic effects also offers advantages in utilizing these molecules for applications in photocatalytic hydrogen production via solar water splitting.[87,88]

To clarify the reasons behind the decrease of the exciton binding energy, we presented the energies of the NFA's low-lying electronic excited states in the Figure 6b. The corresponding spatial distribution of the frontier orbitals associated to these excited sates are illustrated in Figure S4. This figure also indicates the degree of overlap ($\Theta$) between the occupied and unoccupied molecular orbitals with more contribution to these states. It is clear in Figure S4 that the density of HOMO states in the vicinity of the Te (Se) atoms increased relative to the density around the S atoms of unsubstituted NFAs. The higher contribution of the heavier chalcogens to this occupied orbital tends to enhance its spatial delocalization along the central fused-ring core of Y6 (ITIC). The higher HOMO delocalization, which is the basic reason behind the variations of the HOMO energy depicted in Figure 1, is also produced by the extended length of the acceptor central region with the introduction of Se or Te. The HOMO delocalization considerably decrease the fundamental gap of Se or Te of substituted molecules compared to the reduction of the respective excited state energies, resulting in a lowering of $E_{b,S(T)}$, despite the slight increase of $\Theta$.

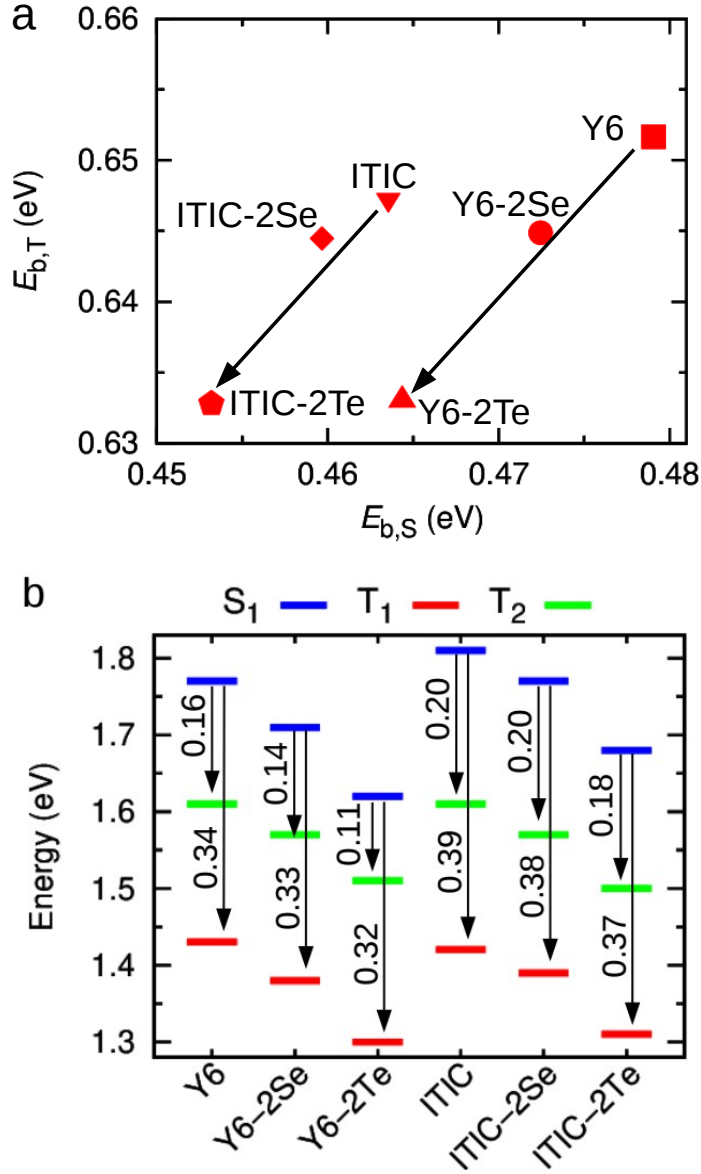

**Figure 6**: a) Binding energy of the singlet exciton and triplet exciton. Inset: the arrows indicates the behavior of energies due to the replacement in Y6 and ITIC of S atoms by Se or Te. b) Excitation energies of $S_1$ and $T_1$ states. Inset: the arrows represent $\Delta E_{ST1} = E_{S1} - E_{T1}$ and $\Delta E_{ST2} = E_{S1} - E_{T2}$.

There are additional important insights from the results in Figure 6b. First, one can see that the singlet ($E_{S1}$) and triplet ($E_{T1}$) energies decreases with Se and Te substitution but the down shift is higher for $E_{S1}$ relative to $E_{T1}$. This effect narrows the singlet-triplet energy gap ($\Delta E_{S1T1} = E_{S1} - E_{T1}$) with selenation or telluration. Since the singlet state tends to be more delocalized than the triplet state in NFAs,[89] $E_{S1}$ is more susceptible than $E_{T1}$ to variations in the conjugation length produced by those substitutions. It is important to clarify that the more localized character of T1 is in terms of the electron-hole separation. In other words, the stronger exchange interaction due to the spatial confinement in shorter conjugated systems destabilizes the singlet relative to the triplet, leading to a size-dependent energy separation.[90]

We note that the value of $\Delta E_{S1T1}$ in Figure 6b for the unsubstituted molecules is in agreement with recent studies that assessed this gap applying different approaches.[23,89] An accurate method to quantify $\Delta E_{S1T1}$ is important because its reduction can potentially improve the photovoltaic performance of NFAs by suppressing the triplet recombination when $E_{T1}$ is shifted above the energy of the CT state.[10,24] Besides that, lower values of $\Delta E_{S1T1}$ were demonstrated to be intrinsically correlated to higher PCE of organic photovoltaics.[91]

Finally, the reduction of $\Delta E_{S1T1}$ can also improve the inter-system crossing (ISC) that tends to increase the population of triplet excitons, which is important to extend the exciton lifetime and diffusion distance of those states. It should be noted, however, that if the triplet excitons are not properly dissociated, there may be increased energy loss and a concomitant decrease in the $V_{OC}$ of OSCs. Please note that the ISC process is also favored by the increasing spin-orbit coupling associated with the incorporation of heavier atoms into the molecular structure.[92] Possible strategies to decrease $\Delta E_{S1T1}$ by chemical modifications of the NFA structure were already envisaged. They included the extension of the fused-ring core[38] and chlorination (or fluorination) of the acceptor end-group.[22] The Te substitution proposed here, however, has the potential advantage of achieving this reduction with a simultaneous stabilization of the fragile bond that links the end-groups with the central core of the NFAs.

Figure 6b revealed that the second lowest triplet excited state (T$_2$) has an energy between S$_1$ and T$_1$ states. For the Y6-like molecules, the reduction of $E_{T2}$ is less pronounced than the reduction of $E_{T1}$ with the Se or Te substitutions. Consequently, the selenation or telluration narrows the $\Delta E_{S1T2}$ gap even more than the $\Delta E_{S1T1}$ gap (for instance, $\Delta E_{S1T2}$=0.11 eV for the Y6-2Te molecule while $\Delta E_{S1T1}$ =0.32 eV). The smaller energy difference of $\Delta E_{S1T2}$ compared to $\Delta E_{S1T1}$ can further facilitates the ISC and enables a effective access to this superior triplet state.[93] The smaller $E_{T2}$ variation compared to $E_{T1}$ may be linked to the increase of the charge transfer character (lower electron exchange energy[94]) of the T$_2$ excited state compared to T$_1$ excited state. This character can be assessed by determining the degree of overlap ($\Theta$) between the occupied and unoccupied molecular orbitals. From our TD-DFT calculations, the orbitals with more contribution to the T2 state are HOMO and LUMO+1 whereas the T$_1$ state is basically derived from a HOMO-LUMO transition. In Figure S4 it is possible to see that $\Theta_{H-L+1}$ is lower than $\Theta_{H-L}$, evidencing the higher charge transfer character of the T$_2$ state compared to the T$_1$ state. Considering that the S$_1\rightarrow$T$_2$ ISC transition can be more favorable than the S$_1\rightarrow$T$_1$ transition due to a lower $\Delta E_{S1T2}$ gap, the population of T$_2$ states pumped from the S1 can quickly relax to the T1 state by internal conversion. This mechanism can help to increase the density of triplet states of NFAs submitted to illumination.

**2.5 Molecular packing**

Molecular dimers were simulated to investigate the packing distance ($d$), the interaction energy between two species (named binding energy, $E_{bind}$) and electronic coupling of LUMO orbitals ($\beta$) which are responsible for electron transport. Because the ITIC-based acceptors have four central side groups perpendicular to the main chain, these molecules interact with each other through the end-groups mainly forming J-type dimers.**[95–98]** In remarkable contrast to ITIC-based NFAs, the Y6-based acceptors have side groups in the backbone plane, which improves molecular packaging, forming both J-type and H-type dimers.[53,89,99] The H-type dimers have a grater molecular superposition than J-type dimer. Therefore, we simulated two configurations for the Y6-based dimers and one configuration for the ITIC-based dimer. Note in Figure 7 that the H-type dimers have a much superior $E_{bind}$ than J-type dimers due to the grater molecular superposition. Comparing $\beta$ between the two dimers type, the difference is even greater, evidencing the consequences resulting from the two forms of molecular packing.

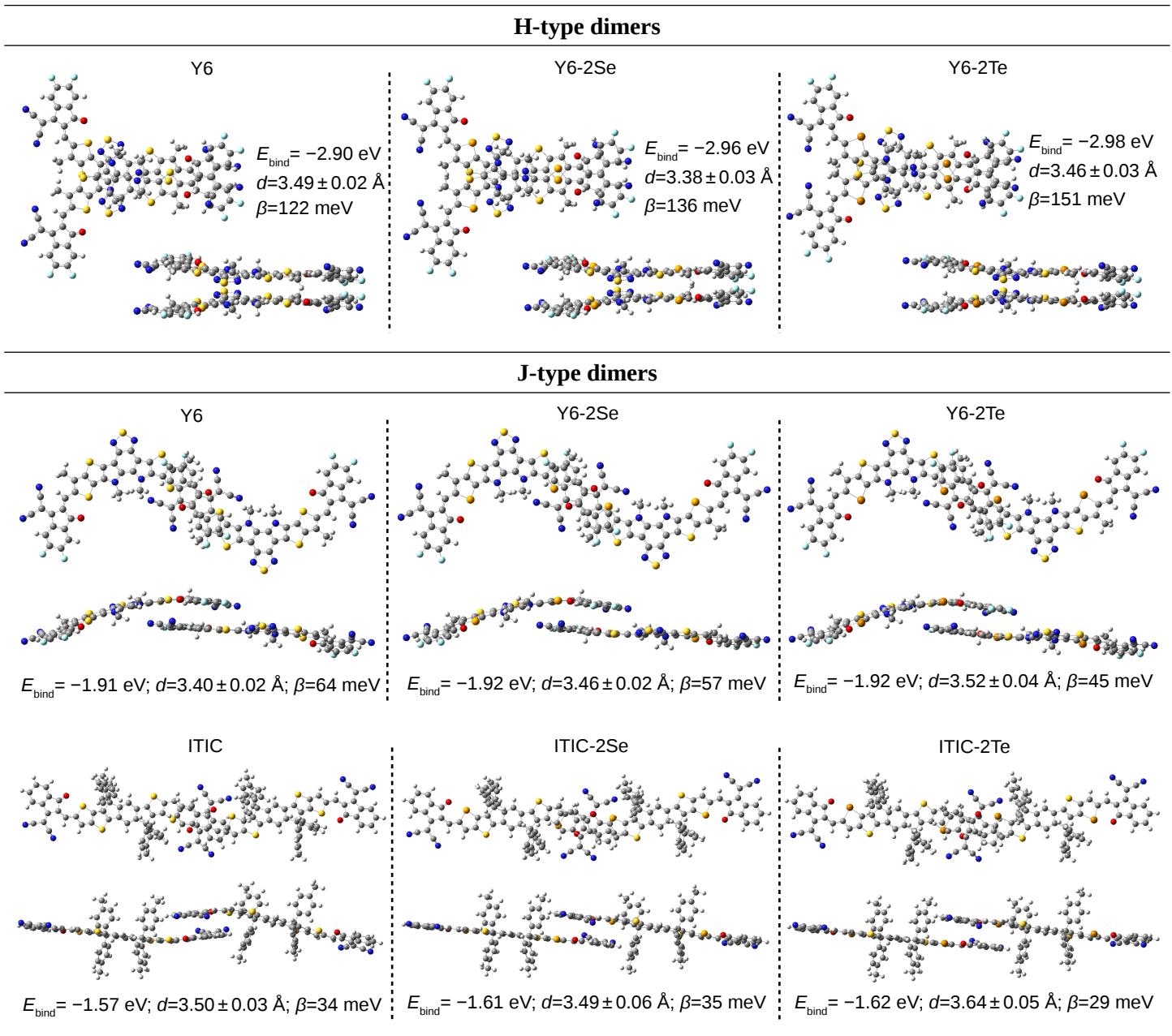

**Figure 7**: Optimized structure of dimers and the calculated values of packing distance ($d$), binding energy ($E_{\text{bind}}$) and electronic coupling of LUMO orbitals ($\beta$). The figure shows a front and side view of the dimers.

Regarding the effects of heavy chalcogenophenes, it is possible to highlight from the dimers of Figure 7 an increase in $E_{\text{bind}}$ following the substitution hierarchy S < Se < Te. The denser packing is due to stronger intermolecular interactions,[53,100] making the molecules with Se and Te atoms less susceptible to photodegradation (better morphological stability).[14,101] In addition, this feature also provides larger crystalline domains, which is important to improve charge transport.[9] Furthermore, it was observed that $\beta$ follows the same trend as $E_{bind}$ for H-type dimers, indicating that intermolecular electron delocalization is also improved with replacement of thiophene by selenophene or tellurophene. This result confirms that strong Te−Te interaction led to strong interchain electronic coupling as it was already observed for tellurophene-based materials.[55] In addition to improve charge transport, electronic delocalization is also important to further decrease the exciton binding energy.[42] On the other hand, $E_{bind}$ and $\beta$ oscillates around very closed values for the J-type dimers without revealing any specific trend. The absence of a systematic variation of $E_{bind}$ and $\beta$ for this kind of dimer may be related to the fact that the heavy chalcogenophenes are located in the central part of them, having little effect on the intermolecular interaction that occurs between the end-groups.

## 3. Conclusions

The main conclusion of this work is that Telluration applied at outer-core positions of A-D-A and A-DA′D-A-type NFA can significantly improve key molecular properties associated to photovoltaic performance and chemical stability. Additionally, it was observed that Selenation, to a somewhat lesser degree, similarly contributes to noteworthy enhancements in pivotal molecular characteristics. The potential improvements include all main processes involved in charge generation: (i) better light harvesting due to a red-shifted light absorption; (ii) easier exciton dissociation promoted by a weaker exciton binding energy; (iii) improved exciton diffusion and charge transport derived from a stronger interchain interaction. Moreover, chalcogen substitution decreases the singlet-triplet energy gap, a critical molecular parameter for reducing voltage loss and triplet recombination. More importantly, structural modifications promoted by chalcogen substitution produces more polarized and rigid molecules that mainly originates from stronger intramolecular secondary bond interactions. This effect occurs between the oxygen in the INCN acceptor group at the edges and the outer chalcogen atom in the central donor core of the NFA (Se⋯O and Te⋯O). The secondary bond effect pulls the INCN moiety in the direction of the molecule's central core. Consequently, there is a shrink of the C–C and C=C bonds that attach the A group to the D core of the molecule, helping stabilize the bonds. Since many mechanisms of NFA degradation involves the break of these bonds, chalcogen substitution may significantly increase the resistance of NFAs to degradation.

Finally, we recognize that tellurium is a scarce chalcogen which can rise objections about the industrial viability of using tellurophene rings to increase the performance and stability of OSC. Yet the very tailored kind of Te incorporation proposed here does not significantly alter the carbon content to a level that compromises the low cost and processability of these new acceptor materials. Hence, the design modification suggested in this work might be a promising technological leap necessary to overcome the commercial barrier imposed by the intrinsic instability of NFA-OSCs.

## Conflicts of interest

The authors declare no conflicts of interest.

## Acknowledgments

The authors acknowledge financial support from LCNano/SisNANO 2.0 (grant 442591/2019-5), INCT - Carbon Nanomaterials and INCT - Materials Informatics. L.B. (grant E-26/202.091/2022 process 277806) and G.C. (grant E-26/200.627/2022 and E-26/210.391/2022 process 271814) are gratefully for financial support from FAPERJ. This study was financed in part by the Coordenação de Aperfeiçoamento de Pessoal de Nível Superior-Brasil (CAPES)-Finance Code 001. The authors also acknowledge the computational support of Núcleo Avançado de Computação de Alto Desempenho (NACAD/COPPE/UFRJ), Sistema Nacional de Processamento de Alto Desempenho (SINAPAD) and Centro Nacional de Processamento de Alto Desempenho em São Paulo (CENAPAD-SP).

## Supporting Information

Computational details, molecular energies, chemical structures, absorption spectrum, density of states, frontier molecular orbitals, and Raman spectra.